\documentclass[pdflatex,sn-basic]{sn-jnl}

\usepackage{graphicx}%
\usepackage{multirow}%
\usepackage{amsmath,amssymb,amsfonts}%
\usepackage{amsthm}%
\usepackage{mathrsfs}%
\usepackage[title]{appendix}%
\usepackage{xcolor}%
\usepackage{textcomp}%
\usepackage{manyfoot}%
\usepackage{booktabs}%
\usepackage{algorithm}%
\usepackage{algorithmicx}%
\usepackage{algpseudocode}%
\usepackage{listings}%

\theoremstyle{thmstyleone}%
\theoremstyle{thmstyletwo}%

\theoremstyle{thmstylethree}%

\begin{document}

\title[MEATVEGSOUP]{Meat, Vegetable, Soup - The First Successful Attempt to Classify Everything}

\author*[1]{\fnm{George} \sur{Weaver}}\email{No Thank You}

\author[]{\fnm{Matthew J.} \sur{Selfridge}}

\author[1]{\fnm{Joseph M.} \sur{Setchfield}}
\author[1]{\fnm{Francesca} \sur{Dresbach}}
\author[1]{\fnm{Vishnu} \sur{Varma}}

\author[1]{\fnm{Juan} \sur{Martinez Garcia}}
\author[1]{\fnm{Ayush} \sur{Moharana}}
\author[1]{\fnm{James} \sur{Keegans}}

\author[1]{\fnm{Lewis J.} \sur{Adams}}

\affil*[1]{\orgdiv{MEATVEGSOUP Collaboration, Arbitrary Classification Research Group}, \orgname{Keele University}, \orgaddress{\city{Keele }, \state{Staffordshire}, \country{UK}}}


\abstract{We present the results of a novel classification scheme for all items, objects, concepts, and crucially -- \textit{things} -- in the known and unknown universe. Our definitions of \textsc{meat, soup} and \textsc{vegetable} are near-exhaustive and represent a new era of scientific discovery within the rapidly-developing field of Arbitrary Classification. While the definitions of \textsc{vegetable} (growing in the ground), \textsc{meat} (growing in an animal) and \textsc{soup} (containing both vegetable and meat) may appear simple at first, we discuss a range of complex cases in which progress is rapidly being made, and provide definitions and clarifications for as many objects as a weekend of typing will allow.}

\keywords{Meat, Vegetable, Soup, AI, Machine Learning}

\maketitle

\section{Introduction}\label{sec1}

Given the recent worldwide debates over the status of the planet/dwarf planet Pluto (wrong, it is in fact a cartoon dog), whether or not something is `AI', and whether Jaffa Cakes are a cake or a pizza \citep{stevance2021, Leembruggen2022}, which have caused untold destruction across communities and groups of scientists and civilians alike, it has become abundantly clear that the field of Arbitrary Classification is one of the most important scientific endeavours of our time.

In fact, several efforts have been made in recent years to create classification and taxonomical schemes \citep{wright2023, Leembruggen2022}. These works, while commendable, lack ambition, and are not applicable outside of the restrictive realms of so-called `astronomy' and so-called `food' respectively.

This re-emphasises the need for a scheme which can provide a complete model of classification for all known and unknown things in the universe, and possibly even beyond. In this paper we have utilised the scheme first introduced by Matthew Selfridge in the late-2010s, with a 3-label model that produces a clear, unarguable output.

\section{The Basic Definition}\label{sec2}

The most simple definitions of the model are as follows:
\begin{itemize}
    \item\textsc{vegetable}: A thing which grows exclusively in the ground.
    \item\textsc{Meat}: A thing which grows exclusively in or from an animal.
    \item\textsc{Soup}: A thing which contains, or is made from, a mixture of both \textsc{vegetable} and \textsc{meat}.
\end{itemize}

\begin{figure}[h]
    \centering
    \includegraphics[width=0.8\linewidth]{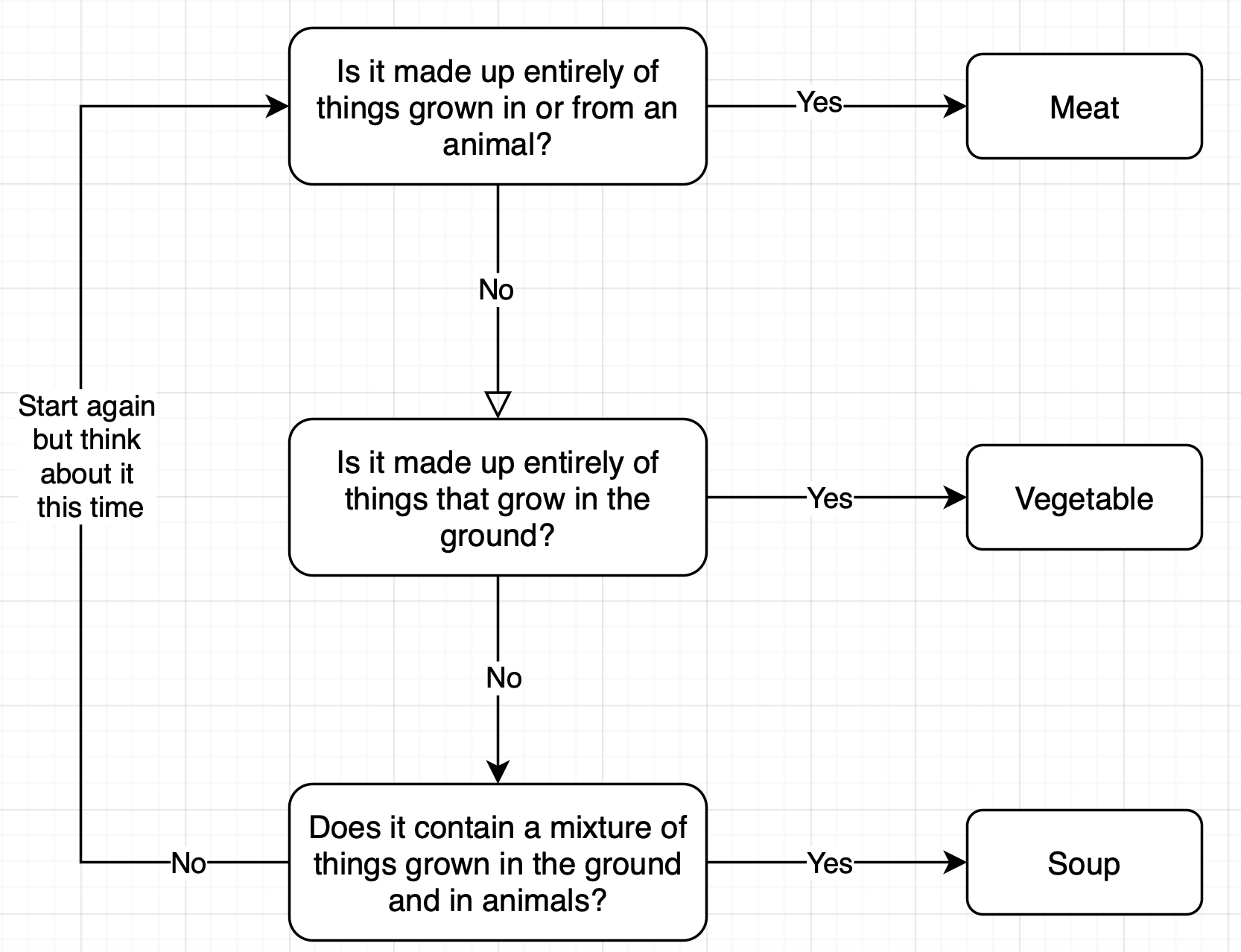}
    \caption{This flowchart defines the process by which the class of a thing is determined.}
    \label{fig:flow}
\end{figure}

This is defined for ease in the flowchart in Fig.\ref{fig:flow}.

The obvious starting points here are with something like grass, which of course grows in the ground, and is thus a \textsc{vegetable}.

Conversely, a rhinoceros grows within an animal, and continues to grow from an animal, making it a \textsc{meat}.

Finally, a delicious chocolate cake contains both \textsc{vegetable} (flour, cocoa, sugar, water) and \textsc{meat} (eggs), meaning it is classified as a \textsc{soup}.

Of course, the greatest difficulty to overcome in this scheme for those new to its power, is the idea of throwing out the old, decrepit, useless definitions of meat, soup, and vegetable that society imposes on us.
In particular, our study (uncited, I promise it's true though) found that over 100\% of newcomers to our scheme were under the ridiculous assumption that \textsc{soup} classification was in any way related to things being in liquid form.
In order to understand this scheme in depth, you must cast off these chains and remember that only the bullet points above are a true definition for each of those 3 classifications.
\\

In case there are more exotic forms of life discovered beyond earth, many have considered it pertinent to define more clearly what we consider an animal.
We, however, feel this is far too pedantic and think it’s obvious what an animal is, so will not be pursuing it further.
\\

As this field evolved in its early stages, it became clear that there are more complex forces at play.
In the following section, we discuss some of the methods and processes to refine our work for more difficult objects, concepts, and things, and the major discoveries already made by our crack team of boffins.

\section{Discoveries}

While the field is still less than a decade old, its discussion across hundreds of lunchtime conversations has led to a vast plethora of discoveries.
Through this section, we detail the most important of these, in chronological order of discovery.

\subsection{The Water Conundrum}

Very early on in the genesis of the model, mere weeks after the infamous `You don't plant bread?', `But you plant \textit{wheat}.' discussion, the conversation arose as to the position of the Earth's water within the categorisation scheme.

After initial discussion, the clarification results in the pioneering understanding that water -- at least within the Earth's water cycle -- is in fact a \textsc{vegetable}.

Let me explain.

The key position here is that water grows in the ground.
On Earth, we see this as the first part of the water cycle, in which springs produce streams that eventually become rivers, seas, and create clouds that form rain.
But even prior to that, water arrived on Earth likely from a comet (which is its own ground), and was created in the lowest density ground in the universe -- space.

\subsection{A True Understanding Of Ground}

Given the groundbreaking discovery that water grows within the ground, it became clear that `\textit{ground}' is a label that requires definition.

We postulate, with the utmost confidence, that the Big Bang led to the first occurrence of ground, which was simply particles and space at that point.

Eventually, the ground began to group together, forming denser areas of ground, which became the stars, planets, moons and asteroids we know and love today.
\\

This leads to two further discoveries that revolutionised the topic:
Firstly, it is clear that the sun, the moon, and all stars and planets and their atmospheres (not including the life living on/in them), are \textsc{vegetable}.
As they are their own ground, and they therefore grow solely in and from the ground, there can be no doubt that this is the case.
Secondly, and most importantly, \textit{ground} itself must be a \textsc{vegetable}, as it is similary grown within the ground of space.

\subsection{The Universe \& The Great Soupification}

This of course moves the debate forward into ideas around the state of the universe as a whole, which in fact turns out to be variable across time.

\begin{figure}[h]
    \centering
    \includegraphics[width=0.8\linewidth]{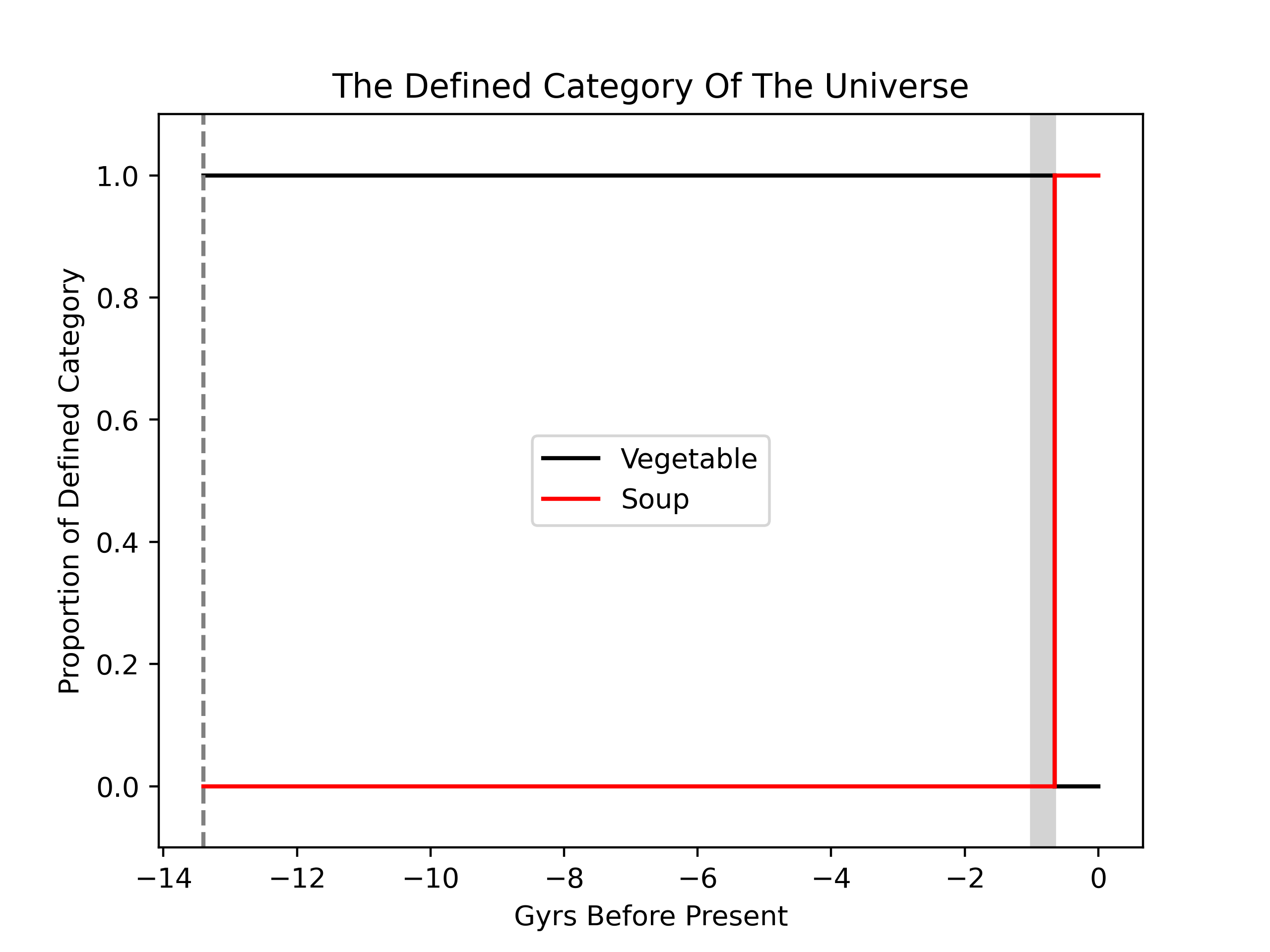}
    \caption{This plot shows how the definition of the Universe as a whole has changed since the Big Bang. The dashed line denotes the Big Bang, and the shaded area denotes the uncertainty in the Great Soupification.}
    \label{fig:universe}
\end{figure}

This is shown in Fig. \ref{fig:universe}, whereby it is clear that, to our knowledge, the Universe has spent the vast majority of its history as a \textsc{vegetable}, before the first animals on Earth evolved.
At this point, instantaneously, the Universe as a whole, now containing both \textsc{meat} and \textsc{vegetable}, became a \textsc{soup}.
We have defined this period, which brainy boffins believe to be approx. 650 Myrs ago \citep{wiki}, as the \textsc{Great Soupification} of the Universe.
\\
This of course backs up earlier discoveries that even the tiniest amount of \textsc{meat} immediately turns a pure \textsc{vegetable} into a \textsc{soup}.
Take the oceans as an example. \textsc{vegetable} for the vast majority of their history, and then \textsc{soup} at the time of the Great Soupification.

The water \textit{in} the oceans remains a \textsc{vegetable}, but the ocean as a whole, with the animals it supports, is a \textsc{soup}.

\subsection{Changes of State}

As you are no doubt wondering, it is in fact possible for these 3 categories to move between one another.

The processes to turn into one of the categories are denoted as \textsc{Soupification}, \textsc{vegification} and \textsc{meatification}.
This discovery allows us to define the points at which, for example, \textsc{vegetable} eaten by an animal, become part of the \textsc{meat} of the animal itself.
\\

For a short while post-consumption, the object of the animal, i.e. its body as a whole becomes a \textsc{soup}, as it now contains \textsc{vegetable} within its digestive system.
The actual animal itself remains a meat, as the \textsc{vegetable} is not yet a part of the animal.
As the animal digests the food, and takes up the nutrients from within it, these nutrients become \textsc{meat}, as they are `grown' in  the animal during the digestion process.
Therefore, any \textsc{vegetable} processed by the animal within digestion, becomes \textsc{meat}.

This process is similar for water intake, as the water taken into the body becomes part of the \textsc{meat} of the fluids used in a multitude of living processes.
\\

Similarly, the process of the vegification of an animal after death occurs as the body becomes part of the ground.
The breaking down of an organism so that its atoms and molecules become part of the Earth is, in effect, a form of `growing in the ground', thus creating \textsc{vegetable} from \textsc{meat}.

\subsection{Concepts}

Put simply, the concept of a thing must always be a \textsc{meat}, as concepts are grown in the brain of animals.
This means, excitingly, that numbers, fears, money, and anything else that is made up, are necessarily \textsc{meat}.
Paradoxically, this also means that if you think of a carrot, it's \textsc{meat}, but if you see one in the mouth of a rabbit, whose name rhymes with Dugs Funny for legal reasons, it's a \textsc{vegetable}.
\\

Additionally, all categorisation schemes are \textsc{meat}, including those cited in this paper.
Some argue that the Meat-Soup-Vegetable scheme is a \textsc{vegetable}, having been formed with the universe as an intrinsic truth of existence.
The key question there is whether this work has discovered, or invented, the scheme we discuss.
The overwhelmingly clear answer to that question is, of course, `yes'.

\section{Future Work}

While we have indeed found a categorisation system that works for everything in the known and unknown universe, there is still significant progress to be made before this project is completed.
A few of the next steps are detailed in this section.

\subsection{The Table Of Everything}

Although a short portion of things are shown in Table \ref{tab1}, it is clear that the pièce de résistance of this work is the generation of a table denoting everything in the universe, and its correct classification.
Work has now begun on this, and we estimate it will be done in about 10 years time, so long as we get funding for it please.

\subsection{Machine Learning Classification Tool}

Following this, we will have sufficient training material to train a classification model using Machine Learning techniques.
This will not provide much use to us, however, as we will have already made the table of everything, as detailed above.

That being said, it doesn't do us any harm to put Machine Learning in this paper for a bit of clout and to be in the zeitgeist.

\subsection{Parallel/Different Universes}

Our classification scheme currently works for the known and unknown universe we live in.
We are uncertain as to whether this applies directly to possible parallel/different universes that may or may not exist.
The possibility of meat stars in other universes remains a significant worry, as does the overwhelming likelihood of soup-based life.

We are therefore exploring expansions to the scheme that can account for these possibilities and hopefully provide some videos for AI content creators on YouTube or preferably Hank Green.

\section{Results \& Tabular Classification}\label{sec5}

In this section you will find a table of as many things as I could fit on one page before \LaTeX's weird formatting ruined it, defined by their categorisation, with notes when required.
\newpage
\begin{table}[h]
\caption{A selection of defined `things'.}\label{tab1}
\begin{tabular}{@{}lll@{}}
\toprule
\textbf{Thing} & \textbf{Classification}  & \textbf{Notes}\\
\toprule
The Sun & Vegetable & \\
\hline
Earth & Soup & \\
\hline
The core of the Earth & Vegetable & \\
\hline
The Moon & Vegetable & Soup for very brief periods\\& & from 1969 - 1972\\
\hline
The Universe & Soup & Vegetable until Great Soupification\\
\hline
The 1998 animated classic `A Bug's Life' & Soup & Production is a soup\\
& & but the projected images\\
&&are vegetable\\
\hline
The bugs in the & Vegetable & Made of CGI\\
1998 animated classic `A Bug's Life' & & \\
\hline
Empty bus & Vegetable & \\
\hline
Bus full of people & Soup & \\
\hline
Sunlight & Vegetable & Grows in the ground (core) \\
\hline
Ocean (Present) & Soup & This includes animals\\
\hline
Water in the Ocean & Vegetable & \\
\hline
Rhinoceros & Meat & Duh \\
\hline
Person riding a rhinoceros with a saddle & Soup & \\
\hline
Person riding a rhinoceros without & Meat & \\ a saddle or clothes \\
\hline
Chicken soup & Soup & \\
\hline
Frozen chicken soup & Soup & \\
\hline
Tomato soup & Vegetable & Soup if cream is included\\
\hline
The idea of tomato soup & Meat & Conceptual \\
\hline
Petrol & Vegetable & Although originally soup \\
& & it's been vegified\\
\hline
WiFi & Vegetable & \\
\hline
Tuesday & Meat & Conceptual \\
\hline
Seven & Meat & Conceptual \\
\hline
Money & Meat & Made up \\
\hline
A steak & Meat & \\
\hline
A seasoned steak & Soup & \\
\hline
A shoe & Vegetable & Unless leather, then soup \\
\hline
A seasoned shoe & Vegetable & As above\\
\hline
Vegetables & Vegetable & \\
\hline
The concept of vegetables & Meat & \\
\hline
ChatGPT & Vegetable & Made of vegetables \\
\hline
Regret & Meat & \\
\hline
Wind & Vegetable & Grows in the ground (atmosphere)\\
\hline
A sneeze & Meat & \\
\botrule
\end{tabular}
\end{table}

\section{Conclusion}\label{sec13}

As is made clear, this system of classification has finally solved the problem of arbitrarily putting everything in the universe and beyond into a small enough number of buckets to feel safe.
We welcome all complimentary responses and comments, and have prepared acceptance speeches for the undoubted incoming Nobel Prizes for physics, literature, and peace.

\backmatter

\bibliography{sn-bibliography}

\begin{thebibliography}{4}
\providecommand{\natexlab}[1]{#1}
\providecommand{\url}[1]{{#1}}
\providecommand{\urlprefix}{URL }
\providecommand{\doi}[1]{\url{https://doi.org/#1}}
\providecommand{\eprint}[2][]{\url{#2}}
 \bibcommenthead

\bibitem[{{ }Wikipedia(2025)}]{wiki}
{ }Wikipedia (2025) {Animal} \urlprefix\url{https://en.wikipedia.org/wiki/Animal#Evolutionary_origin}

\bibitem[{{Leembruggen} and {Martin}(2022)}]{Leembruggen2022}
{Leembruggen} M, {Martin} C (2022) {What's for Lunch? A systematic ordering of foods in the Soup-Salad-Sandwich phase space}. arXiv e-prints arXiv:2203.16580. \doi{10.48550/arXiv.2203.16580}, {\href{https://arxiv.org/abs/2203.16580}{{arXiv:2203.16580}}} {[physics.pop-ph]}

\bibitem[{{Stevance}(2021)}]{stevance2021}
{Stevance} HF (2021) {Using Artificial Intelligence to Shed Light on the Star of Biscuits: The Jaffa Cake}. arXiv e-prints arXiv:2103.16575. \doi{10.48550/arXiv.2103.16575}, {\href{https://arxiv.org/abs/2103.16575}{{arXiv:2103.16575}}} {[astro-ph.IM]}

\bibitem[{{Wright}(2023)}]{wright2023}
{Wright} JT (2023) {A Unified Nomenclature and Taxonomy for Planets, Stars, and Moons}. arXiv e-prints arXiv:2303.18217. \doi{10.48550/arXiv.2303.18217}, {\href{https://arxiv.org/abs/2303.18217}{{arXiv:2303.18217}}} {[astro-ph.EP]}

\end{thebibliography}

\end{document}